\documentclass[conference,10pt]{IEEEtran}

\pdfoutput=1
\usepackage{amsmath}
\usepackage{amssymb,amsmath,amsthm}
\usepackage{graphicx}
\usepackage{amssymb}
\usepackage{cite}
\usepackage{xfrac}
\usepackage{cite}
\usepackage{url}

\usepackage{lipsum}
\newcommand\blfootnote[1]{%
  \begingroup
  \renewcommand\thefootnote{}\footnote{#1}%
  \addtocounter{footnote}{-1}%
  \endgroup
}

\newtheorem{thm}{Theorem}
\newtheorem{lem}{Lemma}
\newtheorem{prop}{Proposition}

\newtheorem{defin}{Definition}
\newenvironment{example}[2][Example]{\begin{trivlist}
\item[\hskip \labelsep {\bfseries #1}\hskip \labelsep {\bfseries #2}]}{\end{trivlist}}
\newenvironment{rem}[1][Remark]{\begin{trivlist}
\item[\hskip \labelsep {\bfseries #1}]}{\end{trivlist}}
\newcommand{\bd}{\begin{defin}}
\newcommand{\ed}{\end{defin}}
\newcommand{\bl}{\begin{lem}}
\newcommand{\el}{\end{lem}}
\newcommand{\br}{\begin{rem}}
\newcommand{\er}{\end{rem}}
\newcommand{\bt}{\begin{thm}}
\newcommand{\et}{\end{thm}}
\newcommand{\be}{\begin{example}}
\newcommand{\ee}{\end{example}}
\newcommand{\bp}{\begin{prop}}
\newcommand{\ep}{\end{prop}}

\newcommand{\bqn}{\begin{eqnarray}}
\newcommand{\eqn}{\end{eqnarray}}

\def\1{\hbox{\rm\rlap {1}\hskip.03in{\rom I}}}
\def\Bbbone{{\rm1\mathchoice{\kern-0.25em}{\kern-0.25em}
{\kern-0.2em}{\kern-0.2em}I}}

\usepackage{cite}
\begin{document}
\title{Robust Detection of Periodic Patterns in Gene Expression Microarray Data using Topological Signal Analysis}

\author{\IEEEauthorblockN{Saba Emrani and Hamid Krim}
\IEEEauthorblockA{Electrical and Computer Engineering Department \\
North Carolina State University\\
Raleigh, NC, US\\
semrani, ahk @ncsu.edu\\}

}
\maketitle
\begin{abstract}

In this paper, we present a new approach for analyzing gene expression data that builds on topological characteristics of time series.
Our goal is to identify cell cycle regulated genes in micro array dataset. We construct a point cloud out of time series using delay coordinate  embeddings. Persistent homology is utilized to analyse the topology of the point cloud for detection of periodicity. This novel technique is accurate and robust to noise, missing data points and varying sampling intervals. Our experiments using Yeast Saccharomyces cerevisiae dataset substantiate the capabilities of the
proposed method.

 \end{abstract}
\begin{IEEEkeywords}
Gene expression, microarrays, topological signal analysis, periodicity
detection, biomedical signal processing

\end{IEEEkeywords}
\section{Introduction}
\blfootnote{*Research supported by National Science Foundation EEC-1160483.} 

Cyclic cellular regulation happens in numerous biological networks due to 
different regulation modes.
Many genes are regulated in a periodic form coincident with the cell cycle. Finding evidence of cyclicity or periodicity in microarray cell cycle data to identify cell-cycle
regulated genes is a challenging goal in gene expression time series analysis due to the following aspects of this dataset.
Mostly, only a small subset of the genes show harmonically changing expression in the cycle and the corresponding time series are remarkably short including a very small number of time points that oscillate in very few number of cycles. Also, periodic gene expression signals are inherently noisy, and dominated by non-cyclic components \cite{Wichert}.
Moreover, gene expression time series are not necessarily evenly sampled and the sampling frequency can vary by time. Finally, there are so many missing values in their expression profiles.
Accordingly, detecting periodic gene expression time series is a very difficult problem. Therefore, development of efficient and accurate techniques for detecting cell cycle regulated genes which is robust to varying sampling rate and missing data points is of great importance.

Different mathematical and statistical methods are applied to micro array time series data in order to determine periodicity \cite{Ahdesma,Johansson, Luan04,Wichert, Lu04, Liew09}.
Yeast Saccharomyces cerevisiae dataset by Spellman et al \cite{Spellman} is a renowned set of gene expression time series utilized in many studies including the ones mentioned above. In Spellman's original paper, cell cycle regulated genes are identified using the combination of a Fourier transform and a correlation algorithm.
A statistical test is utilized in \cite{Wichert} to identify periodically expressed genes based on the g-statistic and the false discovery rate approach is used for multiple testing. The proposed model in this study consists of a simple sinusoidal wave and additive noise while the periodicity detection is performed by searching for peaks in the periodogram.
Moreover, in \cite{Liew09}, singular spectrum analysis, autoregressive based spectral estimation by signal reconstruction, and statistical hypothesis testing are used for periodic signal identification.
In the existing methods, periodic models with a fixed frequency and sampling rate are used and missing data points need to be estimated as part of the pre-processing.
 
Our proposed model for cell cycle regulated genes time series is a sinusoidal function with time varying amplitude, frequency and phase. This model makes our analysis robust to noise since disturbance changes the amplitude and period of the harmonic time series.
 We use time-delay coordinate embedding as a tool to construct point cloud
from  time series. The advantage of this technique is that the point clouds corresponding to the cell cycle regulated genes show cyclic structures even for very short time series.
We can detect the presence of harmonic structures in the data by exploiting topological tools for the analysis of the delay
embedding point clouds. Using topology of the point cloud, we can detect periodicity of the time series in the presence of missing data points.
Persistent homology is used to detect topological holes in the delay embedding point cloud
occurring on account of the periodic structure of the cell cycle regulated genes.

The remainder of the paper is organized as follows: the proposed model and periodicity detection technique is presented in Section 2.
Experimental results are included in Section 3. Finally Section 4 concludes the paper.

\section{Method}
\subsection{The Model} \label{mod}

In this section, we first introduce a model for the general pattern of almost periodic time series. This model is a combination of sinusoidal functions with different frequencies, amplitudes and phases. This representation allows us to detect periodic patterns robustly since the frequency, amplitude and phase of the signal can vary with time. Therefore, we will be able to identify cyclicity in inherently noisy data where these characteristics of the signal are affected by noise. The proposed model in time domain can be expressed as:
 \begin{equation} \label{w}
  y(t)=\sum_{i=1}^n g_i(t),
\end{equation}
where
\begin{equation}\label{g}
  g_i(t) = \left\lbrace  \begin{array}{ccc}
                               y_i(t) &  & t_{i-1}\leq t < t_i,   \\
                               0 &  & \text{otherwise}
                             \end{array}
                             \right.
\end{equation}
and $y_i's, i=1,2,...,n$ are defined as,
\begin{equation}\label{wi}
  y_i(t)=A(t) \sin\left(\omega_i t+\phi_i\right),
  \end{equation}
  where $A(t)$ is a nonzero continuous amplitude function. Also,
to satisfy the continuity of $\omega_i $, the phases $\phi_i$ should conform
to the following condition
  $\phi _{i+1}=\phi _{i}+2\pi t_{i} \left( \omega_i - \omega_{i+1} \right).$
  
\begin{figure}[tb]
      \centering
      \includegraphics[width=7cm]{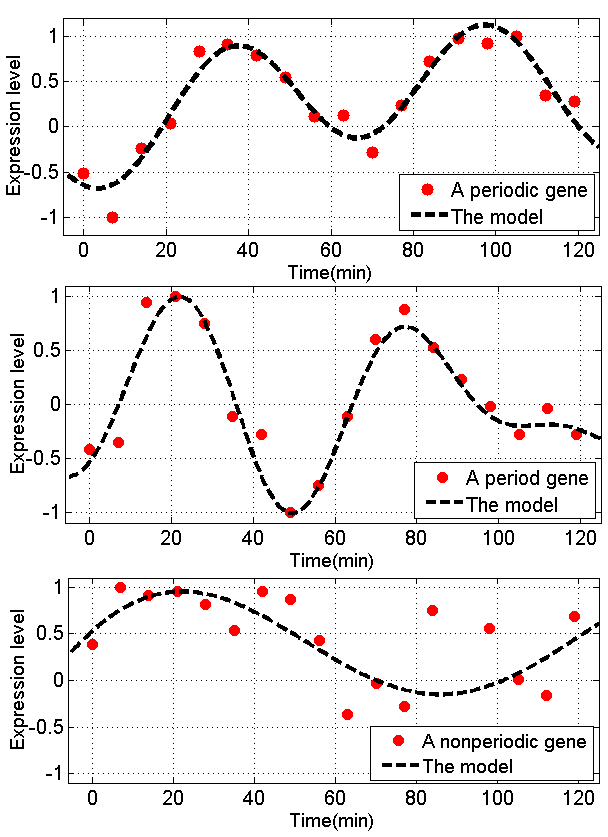}
      \caption{Top: a periodic gene and its corresponding model using fixed frequency and time varying amplitude, middle: a periodic gene and its model using time varying frequency and amplitude, bottom: a non-periodic gene and its model showing a large root mean square error}
      \label{model}
   \end{figure}

  Figure \ref{model} shows two periodic genes and one non-periodic gene with their corresponding models as presented in (\ref{w})-(\ref{wi}). The first plot uses one fixed frequency i.e. $\omega_i=\omega, \forall i$ while amplitude is time varying. The second periodic gene is estimated using both time varying frequency and amplitude. The root mean square error(RMSE) between the periodic genes and their model in Figure \ref{model} are 0.20 and 0.19, respectively. On the other hand if we use the proposed model for estimation of a non-periodic gene, we get a much higher error as RMSE for the one shown in Figure \ref{model} is 0.4877. Note that the expression levels are normalized in $[-1,1]$.
  
According to our experiments using Yeast Saccharomyces cerevisiae dataset, the average RMSE between the periodic genes and our proposed model is $8.50\%$ of the peak to peak value of the signals.
On the other hand, for non-periodic genes RMSE between the model and the signal is very large, with an average of $25.93\%$ of the peak to
peak value of the signals. Thus, the use of this model in the analysis techniques can address the problem
of differentiating between genes with cyclic and non-cyclic patterns quite well.

  

\subsection{Delay Embedding for Point Cloud Construction}
   \begin{figure}[tb]
      \centering
      \includegraphics[width=8cm]{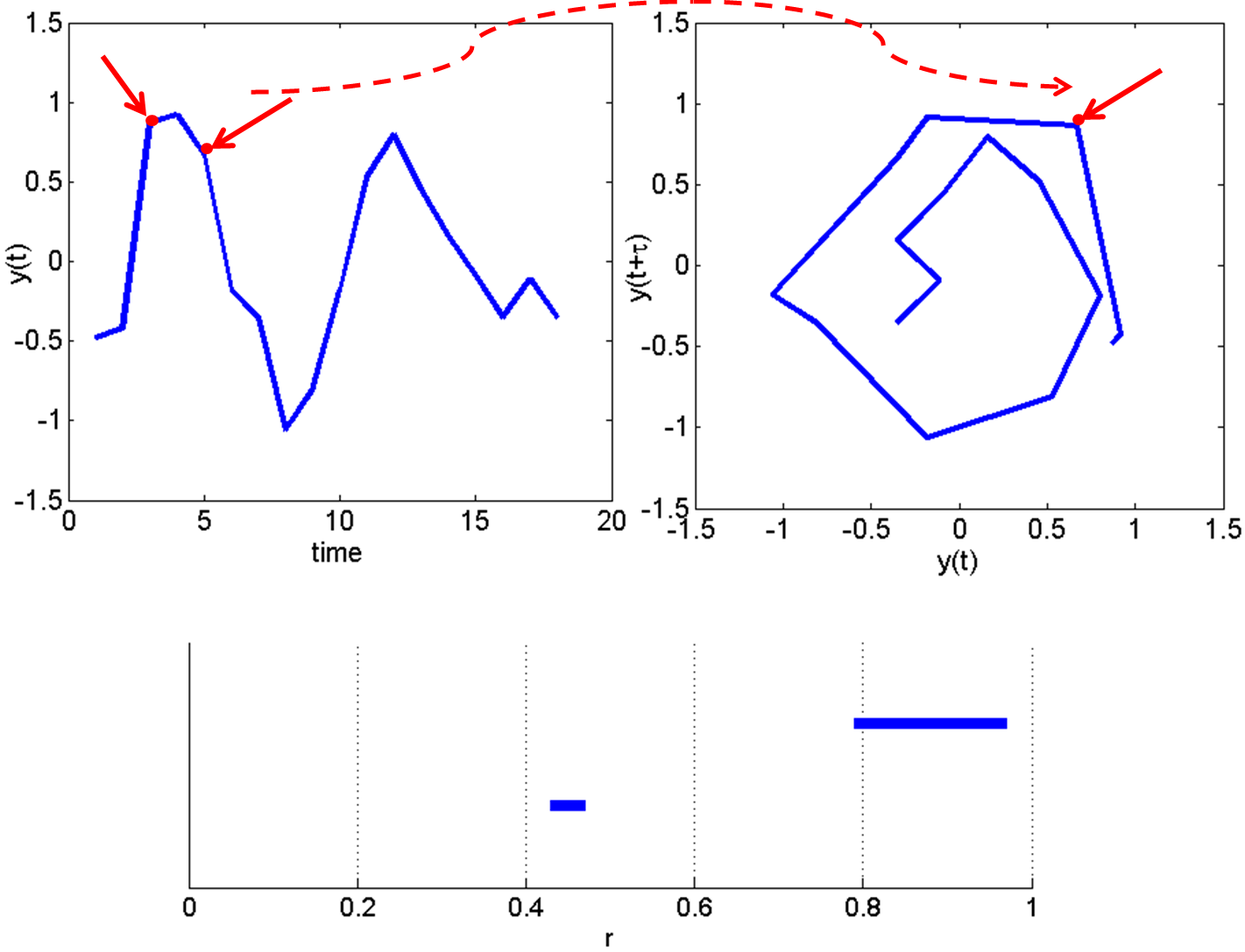}
      \caption{Left: a sample time series, middle: its delay embedding, right: the corresponding barcode}
      \label{dem}
   \end{figure}
The mathematical foundation of delay-coordinate embedding method was proposed by Packard et al. \cite{Pack} and Takens \cite{Takens} to embed a scalar time series into a higher dimensional space.
The 2-dimensional delay-coordinate embedding of time series $\lbrace y_i\rbrace$ can be expressed as

\begin{equation}\label{2d}
  Y(t)=(y(t), y(t+\tau))
\end{equation}
In order for the delay-coordinate components $y(t+\tau_i),i=1,2,..,m-1$ to be independent of each other, the delay time $\tau$ needs to be chosen carefully.
Roughly, if too small, then the the delay embedding is compressed along the identity
line. On the other hand, for too large delays, adjacent components of the delay embedding may become irrelevant \cite{Cas91}. We examined the autocorrelation-like function (ACL) to choose a proper delay time. 
The empirical autocorrelation function of the signal $y(t)$ is calculated as follows
\begin{equation}\label{ac}
R_{yy}(t)=\sum_k y(k+t)y(k)
\end{equation}
Note that for time varying non-stationary signals, a strict autocorrelation function cannot be defined and used for neither establishing periodicity not for detecting the frequencies. However, we use Equation (\ref{ac}) as an autocorrelation-like function strictly for delay selection.
Clearly, peaks in the autocorrelation-like function mark delay times at which the signal is comparatively highly correlated with itself. 
According to experimental results, the appropriate interval for choosing delay time to best obtain informative delay embedding of signal $x(t)$ can be expressed as
\begin{equation}\label{td}
t_{c1}<\tau<t_{c2},
\end{equation}
where $t_{c1}$ and $t_{c2}$ are the first and second critical points of the autocorrelation-like function $R_{yy}(t)$.

%
%
%
We have proved in \cite{me14} that the delay-coordinate embedding $Y(t)$ of $y(t)$ as described in (\ref{w})-(\ref{wi}) is a set of concentric ellipses, with varying
radii and side lengths of the circumscribed squares around them. It is also shown that $Y(t)$ obtained using the appropriate
delay as expressed in (\ref{td}) always has a topological hole. According to the discussions and results presented in \ref{mod}, the model $Y(t)$ approximates the time series of the cyclic genes pretty well. Therefore, there is always a topological hole inside the delay embedding of periodic genes as well. This hole will be used in microarray data in order to detect the periodicity in cell cycle regulated genes.

   \begin{figure}[tb]
      \centering
      \includegraphics[width=4cm]{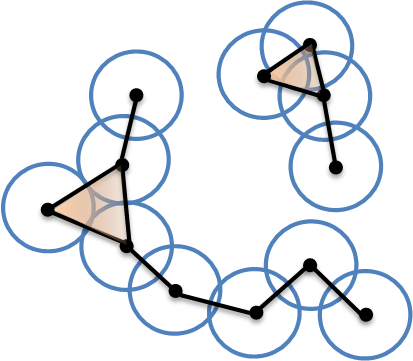}
      \caption{Illustration of persistant homolgy algorithm and barcode construction }
      \label{PH}
   \end{figure}
\subsection{Persistent Homology} \label{PHAlg}

The persistent homology technique is utilized as a tool to identify the hole inside delay embedding of periodic genes using a filtration of the point clouds. This method characterizes the topology of the point clouds using a collection of betti intervals, or persistence intervals which are encapsulated in a structure called persistent barcode. 
Given a point cloud of $n$ points $v_1, v_2,...,v_n$, persistent homology constructs a barcode using the algorithm described below and illustrated in Figure \ref{PH}. We create balls of radius $r$ around each point in the point cloud $v_i$ and denote it by $B_r(v_i)$. The radius $r$ is increased until the balls intersect. For all $0 < i,j < n, i \neq j$, we add a line $[v_1, v_2]$ if and only if $B_r(v_i) \cap B_r(v_i) \neq \emptyset$. While lines are forming and connecting the points, every time the skeleton of a triangle shapes we fill the triangle. To construct the barcode we compute the cycles at each radius value $r$ and represent each cycle(hole) as a point in a 2 dimensional plot where its x-coordinate is $r$. The height of each point is arbitrary but it stays the same for each hole by varying $r$. As we proceed in our filtration sequence by increasing the radius $r$, each Betti interval or bar in the barcode $[r_b,r_d]$ can be associated to a hole, which appears at radius $r_b$ (birth) and ``closes" at $r_d$ (death) \cite{Carlsson08}.

Length of each bar in the barcode is a representation of the size of the corresponding hole. Since the point cloud constructed from the time series of genes with periodic structure always has one dominant hole, the length of the longest bar in the barcode represents that hole. On the other hand, the point cloud of non periodic genes does not contain any dominant hole and only has a few small holes disappearing shortly when $r$ grows. Therefore, barcodes of non-cyclic genes include some short bars that die soon.
Accordingly, we can identify cell cycle regulated genes using a threshold on the length of the longest bar in their persistent barcodes. 
This approach is robust to missing data points since a small subset of the data points in the point cloud preserves the topology of the data set. 
The use of this approach in the micro array data can address the problem of identifying cell cycle regulated genes quite
well.

\section{Results}

The analysis is performed on the yeast Saccharomyces cerevisiae experiment dataset available at http://genome-www.stanford.edu/cellcycle/. This dataset consists of DNA microarrays and samples from yeast cultures synchronized using three different techniques, including $\alpha$ factor arrest (alpha), elutriation synchronization and temperature arrests cdc 15 and cdc28. This dataset contains information about 6178 genes.
The $\alpha$ factor arrest dataset includes 18 time points with a sampling period of 7 mins. The cdc 15 data is sampled at 20 mins for
the first 4 data point, 10 mins for the next 18 time points and again 20 mins for the last two points. The measurements in cdc28 experiment 
are done at intervals of 10 mins from 0 to 160 min with a total of 17 time points. The sampling period of elutriation experiment is 30 mins and it includes a total of 14 points from 0 to 390 min. To date 104 cell cycle regulated genes have been identified in this daatset using traditional techniques \cite{Spellman}.

\begin{figure}[tb]
      \centering
      \includegraphics[width=1\linewidth]{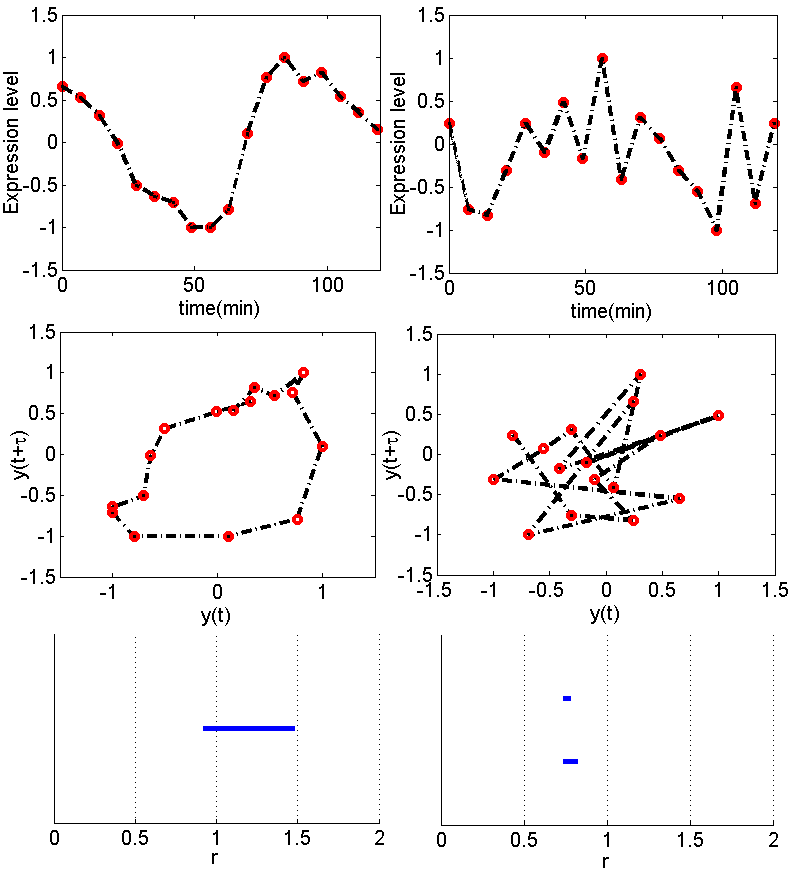}
      \caption{Top row: a periodic (left) and a non-periodic (right) time series. Middle row: their corresponding delay embeddings. Bottom row: their corresponding dimension 1 barcode}
      \label{DEm}
   \end{figure}

\begin{figure}[tb]
      \centering
      \includegraphics[width=0.8\linewidth]{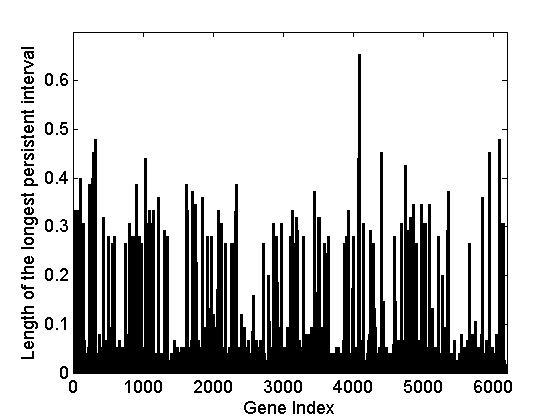}
      \caption{The length of the longest bar in the barcodes of all genes in alpha dataset}
      \label{Thresh}
   \end{figure}

We normalize the amplitudes of all the time series between -1 and 1. In the delay embedding construction, the delays are chosen using Equation (\ref{td}) for the autocorrelation like function. The selected delays for alpha, cdc15, cdc28 and elutriation dataset are 2, 3, 2 and 3 samples, respectively. The delay embedding of all time series are then constructed using equation (\ref{2d}). Persitant homology algorithm as described in \ref{PHAlg} is utilizes using javaplex package for Matlab \cite{javaPlex} to build the corresponding barcodes. Figure \ref{DEm} shows time series of one periodic and one non-periodic gene in alpha dataset in the first row, while their delay embeddings are represented in second row. Clearly, the delay embedding of the cyclic gene contains a dominant hole while the point cloud of the gene with non-periodic pattern does not have any cyclic structure and includes no visible hole. The last row depicts the barcodes showing a long bar for the periodic gene and a few very short bars for the non-cyclic gene data. This experiment is performed on the whole dataset and the length of the longest bar is calculated for each gene. The results for alpha dataset is shown in Figure \ref{Thresh}. Four different thresholds are chosen experimentally and the genes with the length of the longest bar greater than the threshold are considered cell cycle regulated genes. The false negative and false positive numbers for these thresholds (0.16, 0.2, 0.24, 0.28) are tabulated in Table I. Using alpha, elutriation, cdc28 and cdc15 dataset, we were able to identify $85.58\%$, $87.01\%$, $76.33\%$ and $71.60\%$ of the total of 104 cell cycle regulated genes, respectively.


\begin{table}[tb]
\begin{center}
\caption{Experimental results for alpha dataset.} \label{Table1Label}
\begin{tabular}{|c|c|c|}
 \hline
 Thershold & False Negative & False Positive\\
 \hline
 0.16 & 15 ($14.42\%$) & 47 \\
 \hline
 0.2 & 15 ($14.42\%$) & 39 \\
 \hline
 0.24 & 18 ($17.31\%$) & 52 \\
 \hline
 0.28 & 18 ($17.31\%$) & 71 \\
  \hline
\end{tabular}
\end{center}
\end{table}
\section{Conclusion}
In this study, we proposed topological analysis of delay-coordinate embedding to identify cell cycle regulated genes in microarray data. A new model for cyclic gene time series is introduced as a piecewise sinusoidal wave with time varying frequency, amplitude and phase. We described how to apply peristant homology algorithm to delay embedding point cloud of gene expression data. The presented approach were motivated by the challenges in the analysis of microarray cell cycle data including very short time series, many missing data points, intrinsic noise and time varying sampling rates. We applied our methods to the real expression data of Spellman et al. \cite{Spellman} and used periodic genes identified by traditional techniques for comparison of the results.

\bibliographystyle{ieeetr}
\bibliography{Arxive4}

\end{document}